\documentclass{elsarticle}
\usepackage{setspace}
\usepackage{epsfig}
\usepackage{multirow}
\usepackage{subfig}
\usepackage{amssymb}
\usepackage{amsmath}
\usepackage{float}
\usepackage{gensymb}
\usepackage{color}
\usepackage[left]{lineno}
\usepackage[numbers]{natbib}

\begin{document}
\begin{frontmatter}
\title{A Review of Lithium-Ion Battery Models in Techno-economic Analyses of Power Systems}

\author[1]{Anton V. Vykhodtsev}
\ead{anton.vykhodtsev@ucalgary.ca}
\address[1]{Department of Electrical and Software Engineering, University of Calgary, Canada}
\author[2]{Darren Jang}
\ead{Darren.Jang@nrc-cnrc.gc.ca}
\address[2]{Energy, Mining and Environment Research Centre, National Research Council of Canada, Canada}

\author[2]{Qianpu Wang}
\ead{Qianpu.Wang@nrc-cnrc.gc.ca}

\author[1]{Hamidreza Zareipour}
\ead{hzareipo@ucalgary.ca}
\author[1]{William D. Rosehart}
\ead{rosehart@ucalgary.ca}

\begin{abstract}
The penetration of the lithium-ion battery energy storage system (BESS) into the power system environment occurs at a colossal rate worldwide. This is mainly because it is considered as one of the major tools to decarbonize, digitalize, and democratize the electricity grid. The economic viability and technical reliability of projects with batteries require appropriate assessment because of high capital expenditures, deterioration in charging/discharging performance and uncertainty with regulatory policies. Most of the power system economic studies employ a simple power-energy representation coupled with an empirical description of degradation to model the lithium-ion battery.  This approach to modelling may result in violations of the safe operation and misleading estimates of the economic benefits.  Recently, the number of publications on techno-economic analysis of BESS with more details on the lithium-ion battery performance has increased. The aim of this review paper is to explore these publications focused on the grid-scale BESS applications and to discuss the impacts of using more sophisticated modelling approaches. First, an overview of the three most popular battery models is given, followed by a review of the applications of such models. The possible directions of future research of employing detailed battery models in power systems' techno-economic studies are then explored.
\end{abstract}


\end{frontmatter}



\section{Introduction}\label{Introduction}

The number of lithium-ion BESS projects in operation, under construction, and in the planning stage grows steadily around the world due to the improvements of technology \cite{Ziegler2020}, economy of scale \cite{Mauler2021}, bankability \cite{Bonomi2020}, and new regulatory initiatives \cite{FERC841}. It is projected that by 2040 there will be about 1,095GW/2,850GWh of stationary energy storage in operation mostly in the forms of batteries \cite{BNEF2019}. Although the private investor in energy storage for grid applications encounters high capital cost compared with conventional solutions, deteriorating battery characteristics, and revenue risk associated with changing regulatory policies, the economic value of such projects is mostly assessed using a simplistic black-box representation of the battery operation \cite{Walawalkar2007} and an empirical relationship to characterize ageing \cite{Xu2018}. A black-box modelling of a lithium-ion battery is reasonable if the large-scale optimization is solved where additional state variables characterizing a battery would further increase computational complexity. However, short-term operation and long-term planning for the individual storage owner will benefit from detailed models of BESS \cite{Pandzic2019,Jafari2020}. For example, when a lithium-ion battery storage is used to provide multiple services for the electrical grid for better asset utilization and economic benefits for the owner \cite{RockyMountain2015}, the optimal market participation calculated using a simplistic model may lead to the execution of infeasible operations and an erroneous estimate of economic benefits \cite{Taylor2020}. This occurs because a simple battery model can not reflect the physical processes inside a lithium-ion cell, which is the main component of BESS. This can be more pronounced if the operation and planning of BESS for the power system will be performed over multiscale time horizons: control strategy for participating in the electricity and ancillary services markets encompasses hours/minutes/seconds, degradation occurs every second and accumulates over years, and both the planning horizon and the replacement plan should be calculated for years \cite{Sorourifar2020}. The linkage of these timescales will likely be inaccurate if a simple black-box model is used since it operates with power in MW over an hour time interval. The scarce experimental data supports the importance to consider more detailed models for operation \cite{Pandzic2019,Taylor2020} as well as for the long-term performance \cite{Reniers2020}.

In a pioneering work published in 1985, the techno-economic assessment of BESS application was performed by Sobieski \cite{Sobieski1985} where the author compared BESS with combustion turbines for peak shaving capacity expansion and for spinning reserve. Over the years since, the strategic battery operation in various decision-making studies in power systems have been modelled using generic models without a reference to a particular battery technology. Miletic and co-workers \cite{Miletic2020} summarized and structured optimization frameworks and market models for stationary energy storage used in operation and planning problems on the transmission and distribution levels. A standard power-energy model and its various modifications were discussed as well. However, the impact of including additional details into the simple battery model was not assessed. 

The lithium-ion battery community suggests a variety of models with different levels of accuracy and computational complexity for simulation \cite{Ramadesigan2012} and characterization of ageing \cite{Reniers2019}. These models are usually employed in the battery management system (BMS) to predict battery behaviour and to estimate state-of-charge or state-of-health of the battery \cite{Byrne2017}. Until recently the strategic operation of stationary BESS was derived using advanced battery models by only a few researchers\cite{Reniers2020,Cao2020}. The critical review of three models of BESS, namely the energy reservoir model, the charge reservoir model, and the concentration based model, were provided to the power system research community by  Rosewater \textit{et al.} \cite{Rosewater2019} where they used them to calculate the optimal schedule of a BESS for a peak shaving application. The authors outlined the advantages and disadvantages of each model from a computational point of view but they mostly reviewed references outside of typical system-level grid applications of BESS. 

The contribution of the present review paper is to provide a detailed overview of alternative battery models and how they have been used to represent grid-scale BESS in electrical power system studies. In particular, we focus on papers that have integrated transmission-connected BESS into grid operation and planning techno-economic studies. This paper builds on previous works; for example, it extends the work presented in \cite{Rosewater2019} where the models were presented but no discussions of their application in techno-economic optimization problems were conducted. Compared with\cite{Miletic2020}, where different optimization models for BESS operation and planning were summarized, and with \cite{Byrne2017}, where optimization protocols and frameworks for BESS applications were addressed, in this review paper, various applications were examined from the perspective of how the battery was described and represented. The review work \cite{Lawder2014} was focused on the battery models for the BMS and the architecture of BESS and BMS. Their case study only addressed optimal charging scheduling for the combined BESS-photovoltaic generation plant considering physics-based model whereas in this paper a broader range of BESS applications is examined.

The remainder of this paper is structured as follows: the next section gives an overview of BESS models, the third section presents examples of economic studies with BESS providing different services, the fourth section discusses and summarizes the impact of BESS models on the strategic operation and planning. The last section presents the conclusion.

\section{An overview of the lithium-ion battery modelling approaches}\label{An overview of the lithium-ion battery modelling approaches}

A battery is an electrochemical device that is able to store electrical energy in the form of chemical energy and to convert it back to electrical energy when it is needed. Since their invention in 1800 by Alessandro Volta, various battery technologies have emerged; however, in this work, we are focused on the lithium-ion technology. This type of battery was pioneered by Whittingham \cite{Whittingham1976}, significantly improved by Goodenough \cite{Mizushima1980}, and brought to the market by Sony in early the 1990s.  Today, the term battery is often used to refer to the electrochemical storage as a whole system. In fact, each battery consists of a pack of elementary electrochemical units – cells. The way the cells are connected in the battery (in parallel and in series) determines the battery's nameplate ratings.  A cell is a physical place where the conversion occurs and where the electrochemical energy is stored. 

\begin{figure}[t!]
  \centering
  \subfloat[Power-Energy Model]{\includegraphics[trim=7cm 6.5cm 19cm 5.0cm, clip=true,angle=0,width=1.0in]{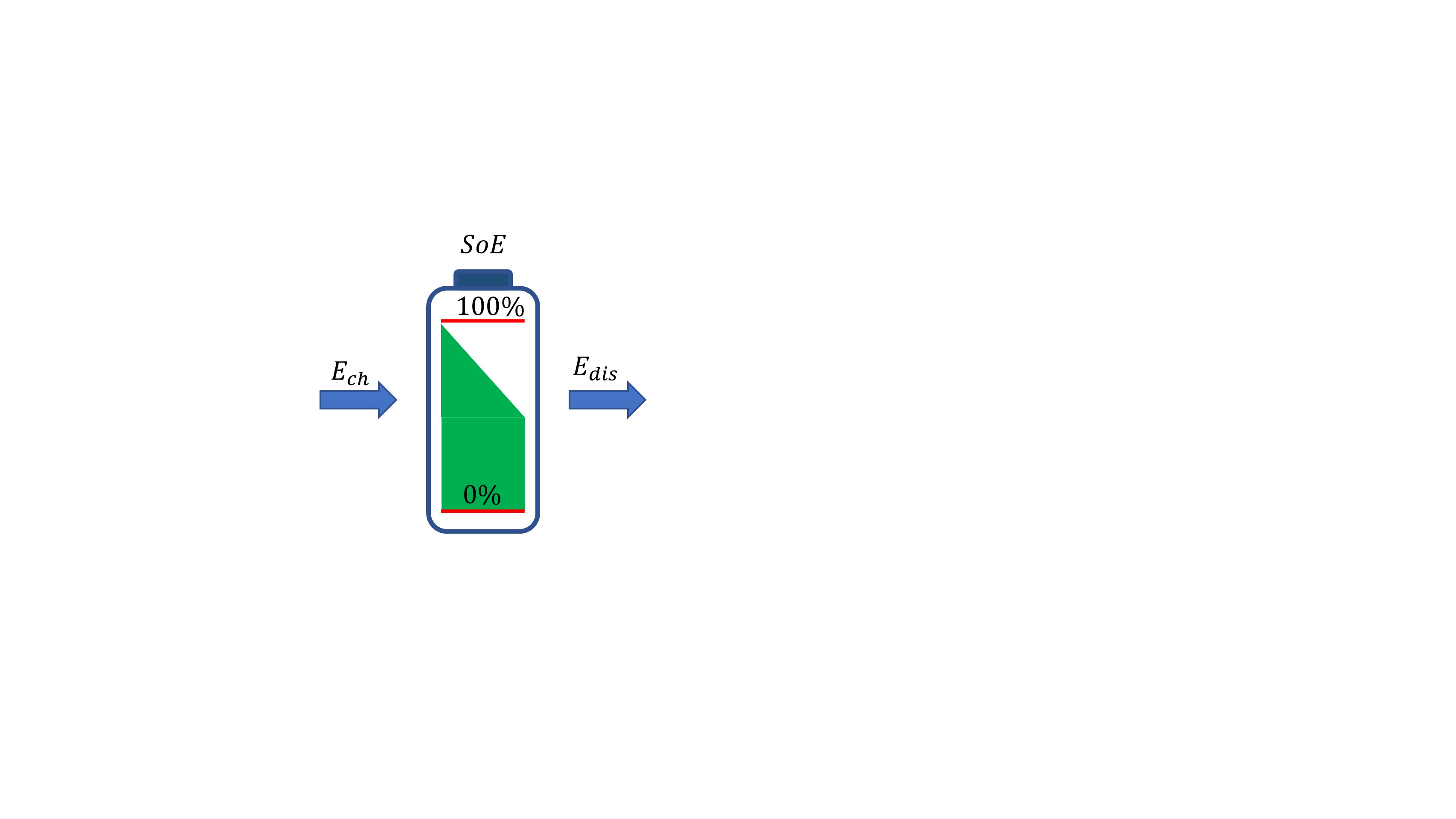}\label{fig:1}}

  \subfloat[Voltage-Current Model]{\includegraphics[trim=8cm 3cm 14cm 5.0cm, clip=true,angle=0,width=1.6in]{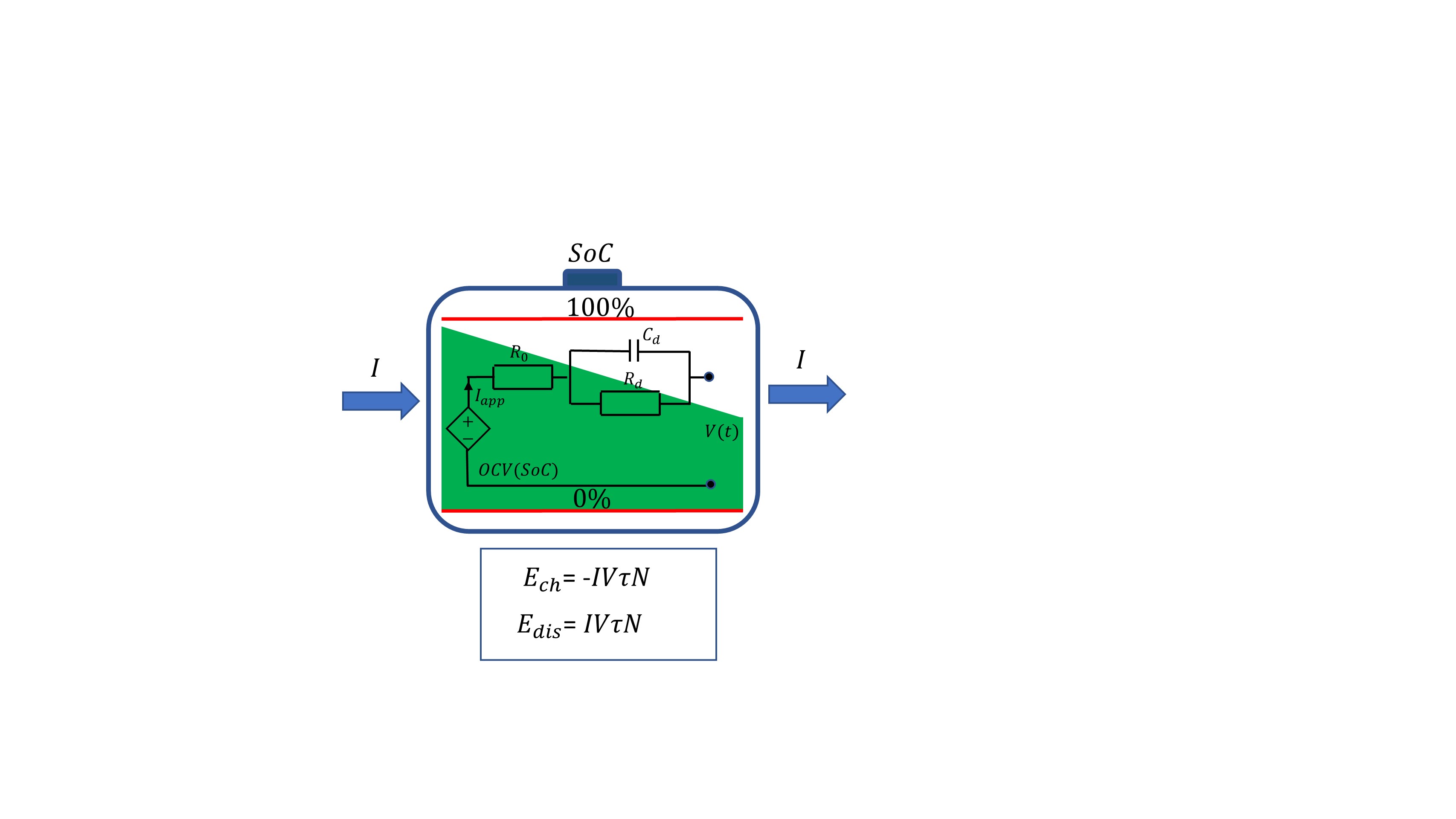}\label{fig:2}}\hspace{0.2cm}
  \subfloat[Concentration-Current Model]{\includegraphics[trim=8cm 3cm 14cm 4.5cm, clip=true,angle=0,width=1.6in]{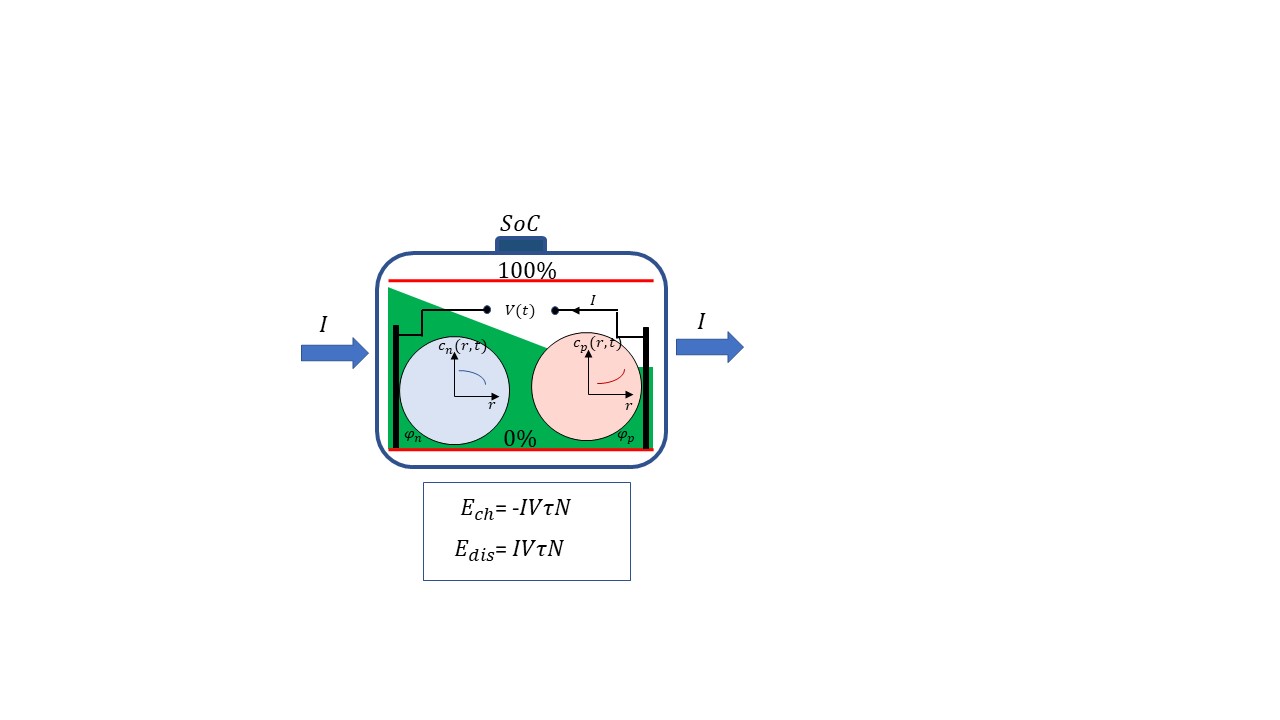}\label{fig:3}}
  \caption{The lithium-ion battery models used in techno-economic analysis of power system.}
\end{figure}

The battery models by the extent of description of the physical processes and corresponding safety constraints can be divided into black-box, phenomenological, and physical models  \cite{Plett2015, Jokar2016}. In the black-box model (Figure \ref{fig:1}), a battery is replaced with a reservoir or a bucket, where energy comes in and comes out. This model does not consider a description of the physical phenomena inside the cell. If the phenomenological model is used (Figure \ref{fig:2}), the battery is replaced with the system that was empirically built to replicate the response of the battery to the control commands \cite{Ramadesigan2012}. The electrochemical process inside the battery and the response of the cell to external factors are accurately described using the physical model (Figure \ref{fig:3}) \cite{Plett2015}. 

The long-term performance of the battery deteriorates with time (calendar ageing) and with the number of  charging/discharging cycles (cycle ageing). There are several mechanical and electrochemical processes that gradually deteriorate either energy capacity, rated charging/discharging power, or both. The ageing of the lithium-ion cells is impacted by environmental conditions, such as high and low temperatures, operating conditions, such as high charging/discharging current and high/low state-of-charge, and even when the battery is at rest \cite{Woody2020}. The degradation of the lithium-ion cell is usually accompanied by loss of lithium inventory or loss of active material of an electrode. The lithium inventory is consumed by various side reactions. The decline of the number of cyclable lithium leads to a decrease in energy capacity and by-products of these parasitic reactions create additional internal resistance of the cell that reduces the available charging/discharging power capacity. Examples of such processes are solid electrolyte interphase (SEI) film growth and lithium plating. The SEI formation is considered as a dominant degradation mechanism \cite{Reniers2019}. The SEI is mostly formed on the surface of the negative electrode during charging since these conditions favour electrochemical decomposition of electrolyte \cite{Arteaga2017}. The lithium ions can also be converted to metallic lithium which is deposited on the electrode through the lithium plating process. The structural changes, such as cracks in the electrode particles or dissolution of material in electrolyte, lead to loss of electrode active material. More detailed information on the degradation mechanisms in the lithium-ion cell can be found in \cite{Woody2020}, \cite{Birkl2017}, and \cite{Reniers2019}.

In this section, the battery models that can be found in power system operation and planning papers are reviewed.

\subsection{Power-Energy Model}\label{Power-Energy Model}

The simplest model of the battery assumes that the battery can be seen as an energy reservoir in which the energy is pumped to store and from which the energy is drawn to consume (Figure \ref{fig:1}). If such a model is used for analysis there is no need to distinguish elementary electrochemical units or the type of electrochemistry within the battery. This is the most popular model to characterize the operation of the battery in techno-economic studies in power systems. It is likely that this modelling approach has come from the mature pumped hydro energy storage modelling that was around for a long time \cite{Bainbridge1966}. The control variables for this model are charging, $ch_{t}$, and discharging, $dis_{t}$ powers whereas state-of-energy, ${SoE}_{t}$, is the only state variable. The state-of-energy indicates the present value of energy (often in MWh in power systems literature) stored in the battery. The BESS is not an ideal system, thus there will be losses during the charging/discharging cycling. The loss in a Power-Energy Model is commonly considered through the introduction of the energy efficiency factor which can be assigned either separately for both charging, $\eta^{ch}$, and discharging, $\eta^{dis}$, operations \cite{Dvorkin2017} or as a round-trip energy efficiency for the whole cycle \cite{Zhao2018,Arteaga2019}. The generic Power-Energy Model assumes fixed energy efficiencies and constant rated charging/discharging power that do not depend on $SoE_{t}$. The evolution of state-of-energy is a core of the Power-Energy Model and the relationship between two consecutive observations of $SoE_{t}$ with a time step $\tau$ between them is expressed as  

\begin{equation}
\label{eq:1}
SoE_{t} = SoE_{t-1}+\tau(\eta^{ch}ch_{t}-\frac{dis_{t}}{\eta^{dis}}).
\end{equation}

The technical aspects of the BESS are considered by enforcing limits on the charging and discharging maximum power and allowing only to store energy until the rated energy capacity is reached. Expression for state-of-energy (\ref{eq:1}) allows the schedule with simultaneous charging and discharging (e.g., that may be realized for electricity  market cases with negative nodal energy prices or for ideal batteries with efficiencies equal to unity \cite{Taylor2015}). In this case, to avoid simultaneous charging and discharging, binary variables are introduced \cite{Go2016}. However, when the cost of BESS operation is included in the objective and efficiencies are less than unity, simultaneous charging and discharging is suboptimal and should not appear in the optimal solution \cite{Zhao2018,Perez2016}. The Power-Energy Model can be updated by adding some features of the lithium-ion cell operation through the functional dependencies of maximum permissible charging/discharging power on state-of-energy as in \cite{Vagropoulos2013,Pandzic2019}, or energy efficiency on state-of-energy and charging/discharging power as in \cite{Sakti2017,Nguyen2019}, or both dependencies as in \cite{Sakti2017,Gonzalez-Castellanos2020,Jafari2020}. A simple Power-Energy Model can also be coupled with degradation description of the battery as result of cycling or calendar ageing. In power system economics studies, degradation is mostly modelled either enforcing operation limits \cite{Mohsenian-Rad2016,Fares2018}, or using the energy throughput model \cite{Wankmuller2017}, or employing the cycle-counting model \cite{Xu2018}. The last two approaches can be included in the optimization framework by assigning the cost of degradation in the cost function or by limiting degradation through the introduction of additional constraints to a Power-Energy Model. References for each method are summarized in Table \ref{tab1}.

In the energy throughput or power-based method, there is a linear dependence between the energy capacity fade and the energy throughput \cite{Wankmuller2017,Arcos-Vargas2020,Fares2018,He2018}. It is assumed that the amount of energy that can be stored and delivered by BESS throughout its lifespan is fixed. A battery is often defined as healthy until it reaches the End-of-life (EoL) state that occurs when the battery has lost 20\% of the original capacity. The framework can be built by incorporating degradation cost in the objective \cite{Wankmuller2017} or by limiting the number of full charging/discharging cycles per day \cite{Mohsenian-Rad2016} or per year \cite{Fares2018}. The energy throughput technique for ageing assessment works properly for one charging/discharging cycle per day \cite{Sorourifar2020}. The cycle-counting degradation model relies on the nonlinear ageing occurred from cycling: cycles with smaller depth-of-discharge (DoD) contribute less into the degradation of the battery \cite{Xu2018}. The cycles are extracted from a state-of-energy profile using the rainflow cycle-counting algorithm \cite{Xu2018a}. Each cycle with a certain DoD is assigned with a fixed amount of degradation to the energy capacity according to the cycle depth ageing stress function that can be obtained from the experimental data. The cycle-counting method is incorporated into the optimization framework by including the cost of degradation into the objective function. This cost is calculated by benchmarking the amount of degradation with the battery replacement cost \cite{Xu2018_Factoring}. Although the cycle-based degradation model is more advanced than the energy throughput method, both techniques do not consider the effect of the average state-of-energy around which the charging/discharging cycle occurs \cite{Maheshwari2020}. Another limitation of the cycle-counting degradation model with rainflow algorithm is that it only has a recursive form. Several approximations suitable for the optimization environment were suggested \cite{Shi2018,Xu2018_Factoring,He2016}. Some authors also employed empirical nonlinear degradation models \cite{Maheshwari2020,Padmanabhan2020} where the cumulative degradation cost function was constructed for different state-of-energy and the current rates for charge/discharge. 

The energy efficiency and maximum power capacity of the BESS also degrades with cycling or when the battery ages over time \cite{Redondo-Iglesias2019}. The effect of ageing on the energy efficiency and maximum charging/discharging power was explored in \cite{He2020} and \cite{Jafari2020}. A linear relationship between the fading capacity and maximum charging/discharging power was assumed in \cite{Jafari2020} and the growth of the cell internal resistance was explored in \cite{He2020}. The coupling of long-term performance with a Power-Energy Model brought problems with computational tractability and is solved either for short optimization horizons (Table \ref{tab1}) or employing sequential approaches \cite{Schneider2020,Maheshwari2020}.

\begin{table}[t!]\centering
\caption{Literature survey on empirical degradation models.}
\begin{tabular}{@{}p{3cm}|p{2cm}|p{2cm}|p{2cm}|p{2cm}@{}}
\hline
\textbf{Decision horizon} &  \textbf{Operational limits} &\textbf{Energy throughput method}& \textbf{Cycle-counting method} &\textbf{Other}\\
\hline

$<$ 1 week & \cite{Wankmuller2017}, \cite{Perez2016},\cite{Mohsenian-Rad2016} &  \cite{Wankmuller2017} &  \cite{Xu2018}, \cite{He2016}, \cite{Shi2018} &  \cite{Maheshwari2020}, \cite{Padmanabhan2020} \\
$<$ 1 year &  \cite{Jafari2020} &  \cite{Fares2018}, \cite{He2018}, \cite{Arcos-Vargas2020}, \cite{Jafari2020} &  \cite{Xu2018_Factoring}, \cite{Schneider2020} &   \\
$<$ 20 year &  & \cite{Sorourifar2020} &  &   \\
  
\hline
\end{tabular}
\label{tab1}
\end{table} 

The incorporation of a generic Power-Energy Model within power system optimization frameworks usually leads to a linear programming problem \cite{Fares2018} or a linear mixed integer programming problem \cite{Sakti2017} that can be easily solved with standard commercial solvers.

\subsection{Voltage-Current Model}\label{Voltage-Current Model}

The charging/discharging schedule calculated using a Power-Energy Model may lead to the operation of the battery out of permissible range for current and voltage \cite{Pandzic2019}. If this occurs the ``optimized" operation will be corrected by BMS \cite{Byrne2017}. This will lead to a deviation from estimated financial benefits and grid service commitment in power systems studies. The formulation of the battery model can be improved if some details of the battery operation are incorporated. A phenomenological model, such as the equivalent-circuit model, is designed to replicate the battery's charging/discharging performance and is usually used in BMS \cite{Geng2020}. The description of BESS operation based on the electric circuit presents an attractive option to model the cell using the Kirchhoff equations. 

The equivalent-circuit model by its nature does not model the dynamics of the internal processes inside the cell but characterizes the measurable response of the cell to the external influence. The charging/discharging performance curves show how the voltage, which is the state variable, across the cell changes with the current, which is the control variable, flowing through the cell while charging/discharging. With regard to the decision variables when an equivalent-circuit model is employed for the optimal control in the power system, the model can be referred to as a Voltage-Current Model. The impedance parameters of the Voltage-Current Model are obtained from fitting an experimental data or the manufacturer's specification to the governing equation of the suggested circuit model. 

There are various configurations of an electric circuit for a Voltage-Current Model \cite{Hu2012}; the choice depends on the accuracy requirement, the level of tolerance to the computational complexity, and the cell chemistry. The first-order approximation consists of two elements: voltage source and resistance \cite{Plett2015}. A more advanced Voltage-Current Model usually consists of a set of resistors and capacitors in series or parallel, current sources, and special nonlinear elements. The simple electric circuit that captures the main processes in the cell is shown in Figure \ref{fig:2}. This model was derived using \cite{Plett2015} and is based on empirical observations. The variable voltage source changes its output as a function of state-of-charge. This voltage is sometimes referred to as an open-circuit voltage \cite{Plett2015} and is supplied by the electrode chemistry.  The nonideality of the cell is modelled through resistor $R_{0}$. This resistor also ensures that the output voltage drops relatively to the open-circuit voltage when the load is connected, and increases above the open-circuit voltage during charging operation. The lithium ions do not immediately stop flowing when the charging/discharging is interrupted. To model this diffusion process, a resistor, $R_{d}$, and capacitor, $C_{d}$, in parallel are employed. The nonlinear elements are used to simulate the so-called hysteresis effect when an open-circuit voltage reaches different values for the same state-of-charge as a consequence of thermodynamic hysteresis or mechanical hysteresis \cite{Ovejas2019}. The latter is a consequence of mechanical stress on electrodes from lithiation and delithiation and the former originates from the variation of the lithium intercalation rates between particles of active electrode material.

The control variable for the Voltage-Current Model is current through the cell $I_{t}$. It is chosen to be positive during the discharge for consistency of the description. The state variables, in this model, are the state-of-charge measured in Ah, the operating voltage  $V_{t}$, and the diffusion voltage $V^{d}_{t}$, i.e., the voltage across a capacitor. The evolution of the state-of-charge is given as:

\begin{equation}
\label{eq:5}
SoC_{t} = SoC_{t-1}-\eta_{c} I_{t}\tau,
\end{equation}
where $\tau$ corresponds to the time step between two estimates of $SoC_{t}$, and $\eta_{c}$ stands for the coulombic efficiency that reflects how much charge is lost during the charging/discharging cycle.

Using Kirchhoff's Law for voltage, the following expression can be derived to relate voltages in the circuit \cite{Plett2015}:

\begin{equation}
\label{eq:6}
OCV_{t}(SoC_{t}) = I_{t}R+V^{d}_{t}+V_{t}.
\end{equation}

Using Kirchhoff's Law for current, the relationship for currents in a parallel RC branch is given as \cite{Plett2015}:

\begin{equation}
\label{eq:7}
I_{t} = \frac{V^{d}_{t}}{R_{d}}+C_{d}\frac{dV^{d}_{t}}{dt}.
\end{equation}

If the derivative $\frac{dV^{d}_{t}}{dt}$ is approximated using finite differences, the diffusion voltage ${V^{d}}_t$ can be calculated as:

\begin{equation}
\label{eq:8}
V^{d}_{t} = \frac{R_{d}C_{d}}{\tau+R_{d}C_{d}}V^{d}_{t-1}+\frac{\tau R_{d}}{\tau+R_{d}C_{d}}I_{t}.
\end{equation}

The Voltage-Current Model is formulated using the single cell perspective which assumes that all cells within the battery show the same performance \cite{Reniers2018}. Although the battery balancing scheme provided by BMS is intended to maintain variation between the cells of the battery pack at minimum \cite{Omariba2019}, in general, this statement requires additional investigation \cite{Fantham2020,Rosewater2019}. DC Voltage and current are not typical variables in power system economic studies, but both of these can be employed  naturally to find the power supplied or consumed by a BESS that consists of $N$ lithium-ion cells in series and in parallel, as follows:

\begin{equation}
\label{eq:9}
P_{t} = NI_{t}V_{t}.
\end{equation}

The Voltage-Current Model allows for the restrictions specified by the producer of the lithium-ion cell in the box constraint on current and voltage, as follows:

\begin{equation}
\label{eq:11}
V^{Min}\leq V_{t}\leq V^{Max}
\end{equation}
\begin{equation}
\label{eq:12}
-I^{MaxCh}\leq I_{t} \leq I^{MaxDis},
\end{equation}
where $V^{Min}$ and $V^{Max}$ identify operational limits for voltage, and  $I^{MaxCh}$ and $I^{MaxDis}$ are maximum absolute values of continuous charging and discharging currents. The range of $SoC_{t}$ is limited from one side by nominal capacity $Q^{Max}$ in [Ah], as follows:

\begin{equation}
\label{eq:10}
0\leq SoC_{t} \leq Q^{Max}
\end{equation}

The model presented above is discussed in more detail in \cite{Plett2015} and was used in the optimization framework by \cite{Reniers2018}. In contrast, Taylor \textit{et al.} \cite{Taylor2020} used a simpler electric circuit composed of the voltage source and a resistor in their work to derive a strategic operation of BESS. Other electric circuits that can be used in optimization frameworks can be taken from \cite{Hu2012}. The examples of the other equivalent-circuit models derived from the physics-based models can be found in \cite{Berrueta2018,Varini2019,Li2019}. The degradation can be included in the Voltage-Current Model using the energy throughput as in \cite{Reniers2018,Li2019}. The Voltage-Current model can also be built without a reference to the underlying electric circuit and operates using empirical relationships. For instance in \cite{Aaslid2020}, the voltage of the lithium-ion cell, as a function of charging/discharging current and state-of-charge, was constructed using bi-variate cubic splines.

The incorporation of a Voltage-Current Model, which is governed by equations (\ref{eq:5}),(\ref{eq:6}),(\ref{eq:8})-(\ref{eq:12}), into the optimization framework leads to a nonlinear programming problem because of a nonlinear relationship between open-circuit voltage and state-of-charge. The final problem can be solved with various off-the-shelf nonlinear solvers such as IPOPT \cite{Wachter2006} or using linearization techniques combined with the commercial linear solvers.

\subsection{Concentration-Current Model}\label{Concentration-Current Model}

Despite having a number of advantages over the Power-Energy Model, the Voltage-Current Model does not provide information about the physical process inside the battery and can produce errors if it is used outside of the operating conditions for which it was empirically built \cite{Plett2015}. In contrast, the physics-based electrochemical model of a lithium-ion cell can achieve better accuracy \cite{Reniers2018}. The enhanced model of a lithium-ion cell is able to characterize the transport of charge carriers, interfacial reactions, thermal effects, and their mutual effects on each other \cite{Plett2015}. The most rigorous model is too complex for the optimization framework because the model contains coupled partial differential equations and nonlinear algebraic expressions \cite{Pandzic2019}. 

The trade-off between accuracy and possibility to implement a more advanced model in the optimization framework can be found in the single particle model of the lithium-ion cell (Figure \ref{fig:3}). The single particle model limits its consideration to the physical principles such as the transport of lithium in the active material of electrodes and the kinetics of the lithium intercalation/deintercalation reactions \cite{Bizeray2018}. The model originates from the porous electrode theory \cite{Newman1975}, which is used to quantify electrode processes within the porous electrodes. The single particle model is built on several assumptions \cite{Ning2004}. First, the active material of both electrodes is composed of uniform spherical electrode particles, which all have an equal radius $R^i$ (the superscript $i$ is replaced by $p$ for positive electrode and it is changed to $n$ for negative electrode). A single electrode particle is employed to simulate the transport of lithium in the active material of the electrode. Second, the concentration of the lithium ions in the electrolyte is assumed to be uniform and constant. Finally, the rate of electrode reaction at the electrode/electrolyte interface does not change from one electrode particle to another. The second and third assumptions are valid in cases of low to medium current through the cell \cite{Bizeray2018} when the impact of the electrolyte potential is negligible \cite{Schmidt2010}.  

The movement of lithium under the concentration gradient in the electrode particle with radius $R^i$ is described by a one-dimensional parabolic partial differential equation in spherical coordinates, as follows \cite{Ning2004}:
\begin{equation}
\label{eq:13}
\frac{\partial c^i}{\partial t}=\frac{D^i}{{r^i}^2}\frac{\partial}{\partial r^i}({r^i}^2\frac{\partial c^i}{\partial r^i}),
\end{equation}
where $c^i$ stands for the concentration of lithium atoms in the electrode particle, $r^i$ is a radial coordinate, and $D^i$ is the diffusion coefficient of lithium in the electrode active material. The homogeneous Neumann boundary condition is applied at the center of the electrode particle to conserve symmetry, as follows \cite{Bizeray2018}: 

\begin{equation}
\label{eq:14}
(D^i\frac{\partial c^i}{\partial t} )_{r^i=0}=0
\end{equation}

The molar flux of lithium ions, i.e., the reaction rate of the deintercalation/intercalation process, $J^i$ on the surface of the electrode sets the Neumann boundary condition for the diffusion equation, as follows \cite{Guo2011}: 

\begin{equation}
\label{eq:15}
(D^i\frac{\partial c^i}{\partial t} )_{r^i=R^i}=-J^i.
\end{equation}

The initial condition is defined as \cite{Ning2004}:

\begin{equation}
\label{eq:16}
(c^i(r^i,t))_{t=0}=c^i_0(r^i),
\end{equation}
where $c^i_0(r^i)$ stands for the initial concentration of lithium in the electrode. This concentration depends on the initial state-of-charge of the lithium-ion cell. The diffusion equation with the boundary conditions presents a challenge for incorporating it as a constraint in the optimization framework. The ``optimization-wise" formulation can be derived through the finite differences of the original partial differential equations \cite{Plett2015} or their approximations based on the ordinary differential equations \cite{Subramanian2001, Wang1998}, or using the Chebyshev collocation method \cite{Bizeray2018}. 

The electrode reaction is characterized by the Butler-Volmer kinetics equation \cite{Guo2011}. This expression describes the rate of the electrode reaction, i.e., the molar flux of lithium ions, at which lithium ion consumes electron and converts to neutral atom inside the electrode or vice versa and is expressed as:

\begin{equation}
\label{eq:20}
J^i=2j^i_0\sinh(\frac{F\eta^i}{2RT}),
\end{equation}
where $F$ is the Faraday constant, $R$ is the gas constant, and $T$ is temperature. The activation overpotential $\eta^i$ is responsible for driving the current that was generated at the electrode during the lithium intercalation/deintercalation process. The exchange current density $j_0^i$ represents the oxidation and reduction currents without external impact and is given as:
\begin{equation}
\label{eq:21}
j^{i}_{0}=k^{i}F \sqrt{(c^{Max,i}-c^{\textrm{surf},i})c^{\textrm{surf},i}c^{\textrm{el}}},
\end{equation}
where $k^{i}$ denotes the reaction rate constant, $c^{\textrm{surf},i}$ stands for the lithium concentration at the surface of the electrode particle, $c^{Max,i}$ is the maximum concentration of lithium atoms in the electrode particle, and $c^{\textrm{el}}$ is the electrolyte concentration (constant for single particle model). 
	
Another parameter to characterize the lithium-ion cell is the so-called equilibrium potential or the open-circuit potential of the electrode that shows how the Gibbs free energy changes when lithium ions enter/leave the electrode \cite{Liu2016}. The functional dependence between the concentration of lithium on the surface of the electrode, $c^{\textrm{surf},i}$, and open-circuit potential, $OCP^i$, is determined experimentally for each type of electrode chemistry when there is no current flowing through the cell (equilibrium state). When the cell is charging or discharging, the potential of the electrode deviates from the open-circuit potential, and is known as the solid-phase potential, $\phi^i$, and is expressed as:
 
\begin{equation}
\label{eq:23}
\phi^i=\eta^i+OCP^i(c^{\textrm{surf},i})+I^iZ^i,
\end{equation}
where $I^i$ is the total current through electrode and $Z^i$ stands for the resistance of the film on the electrode surface. Finally, the single particle model relates the applied charging/discharging current with the rate of the electrode reaction through equation (\ref{eq:24}) for the positive electrode and equation (\ref{eq:25}) for the negative electrode respectively, as follows:

\begin{equation}
\label{eq:24}
J^p=-\frac{IR^p}{3\nu^p\varepsilon^pF}
\end{equation}
\begin{equation}
\label{eq:25}
J^n=\frac{IR^n}{3\nu^n\varepsilon^nF},
\end{equation}
where $\varepsilon^p$ and $\varepsilon^n$ denote the volume fraction of active material in the corresponding electrode, $nu^p$ and $nu^n$ are the volumes of each electrode.

Finally, the voltage of the lithium-ion cell is given as:
\begin{equation}
\label{eq:27}
V_{t}=\phi^p-\phi^n.
\end{equation}

From an optimization perspective the single particle model \cite{Reniers2018,Cao2020} introduces current and concentration as decision variables, thus it will be referred to in this work as the Concentration-Current Model. Similar to a Voltage-Current Model, the Concentration-Current Model represents only one lithium-ion cell and the projection of this model on the whole battery requires an assumption that all cells behave identically. The supplied and consumed power by a battery composed of $N$ lithium ion cell is derived through applied current and the voltage across the cell:
\begin{equation}
\label{eq:28}
P=NI_tV_{t}
\end{equation}

The Concentration-Current Model allows introducing specifications of the cell in the box constraint form, as follows:

\label{eq:30}
\begin{equation}
V^{Min}\leq V_{t}\leq V^{Max}
\end{equation}
\begin{equation}
\label{eq:31}
-I^{MaxCh}\leq I_{t} \leq I^{MaxDis},
\end{equation}
where $V^{min}$ and $V^{max}$ identify operational limits for voltage, and  $I_{ch}^{max}$ and $I_{dis}^{max}$ are maximum continuous charging and discharging currents.

The capacity of the cell is constrained in the Concentration-Current Model through the lithium concentration in both electrodes:
\begin{equation}
\label{eq:29}
c^{i, Min}\leq c_t^{i} \leq c^{i, Max}
\end{equation}  
where $c^{Min,i}$ and $c^{Max,i}$ are limits for lithium concentration in an electrode.

The state-of-charge is not employed in the Concentration-Current Model as a state variable, but it can be used for comparison with other battery models. It can be derived from the instantaneous lithium concentration in one of the electrodes, for example in the negative electrode \cite{Plett2015}, it is given as: 
\begin{equation}
\label{eq:35}
SoC=\frac{ c_t^{\textrm{surf},n}-c^{Min,n}}{c^{Max,n}-c^{Min,n}}Q,
\end{equation}
where $Q$ is the rated capacity of the lithium-ion cell.

When the Concentration-Current Model (\ref{eq:13})-(\ref{eq:29}) is a part of the optimization framework, the whole decision-making problem is a nonlinear programming problem. This problem can be solved using a nonlinear commercial solver such as IPOPT \cite{Wachter2006}.  

The Concentration-Current Model can be naturally updated to include the physical description of the degradation \cite{Plett2015}. The growth of SEI is selected as a principal contributor to the degradation process \cite{Pinsona2013}. The SEI mathematical model employed here was taken from \cite{Ramadass2004,Ning2004}. The rate of the side reaction responsible for the formation of SEI is governed by the Tafel equation, as follows: 

\begin{equation}
\label{eq:36}
J^{\textrm{sei}}=\frac{j^{\textrm{sei}}_0}{F}\exp(\frac{F}{2RT}\eta^{\textrm{sei}}),
\end{equation}
where $\eta^{\textrm{sei}}$ is the overpotential of the side reaction and $j^{\textrm{sei}}_0$ stands for the exchange current for the side reaction. The overpotential  $\eta^{\textrm{sei}}$ is expressed as:
\begin{equation}
\label{eq:37}
\eta^{\textrm{sei}}=\phi^n-OCP^{\textrm{sei}}-I^nZ^n,
\end{equation}
where $OCP^{\textrm{sei}}$ is the open-circuit potential of the side reaction. The total current through the negative electrode is composed of the intercalation/deintercalation current and the side reaction current density and is given as:

\begin{equation}
\label{eq:38}
J^{\textrm{sei}}+J^{\textrm{Li},n}=J^n
\end{equation}

The resistance of the SEI film on the surface of the negative electrode increases as SEI forms \cite{Ning2004} and is expressed as:
\begin{equation}
\label{eq:39}
Z^n=Z^{0,n}+\frac{\delta^{\textrm{sei}}(t)}{\kappa},
\end{equation}
where $\delta^{\textrm{sei}}$ refers to the thickness of SEI film and $\kappa$ is the ionic conductivity of SEI. It can be reasonably assumed that the rate of SEI growth is proportional to the rate of the side reaction \cite{Ramadass2004} and is given as:
\begin{equation}
\label{eq:40}
\frac{d\delta^{\textrm{sei}}(t)}{dt}=-\frac{J^{\textrm{sei}}M}{\rho},
\end{equation}
where $M$ denotes the molar mass of SEI, $\rho $  is the density of SEI. Equation (\ref{eq:40}) can be converted to the discretized version when including in the optimization framework. The increase in $Z^n$ will lead to the decline in charging/discharging power. Finally, the loss in the lithium inventory due to the SEI formation during charging can be estimated as:

\begin{equation}
\label{eq:42}
C_{loss}=-\int_{t_1}^{t_2}\frac{3\varepsilon^p\nu^pJ^{\textrm{sei}}}{R^p}dt,
\end{equation}
where  $[t_1,t_2]$ is the time interval during which the cell was charging.

Equations (\ref{eq:36})-(\ref{eq:42}) can be included into nonlinear optimization frameworks. The single particle model can be improved by adding description of the lithium-ion transport in electrolyte as in \cite{Perez2016_2} to derive the optimal charging protocol or as in \cite{Gailani2020} for assessing revenue in the capacity market under given operation schedule.

\subsection{Summary}\label{Summary}

The reviewed battery models are found to be employed in the decision-making problems that include stationary lithium-ion battery storage for power system level applications; those applications are discussed in the next section. These three models can be converted to one another: the Power-Energy Model can be seen as the Voltage-Current Model with constant voltage that is equal to the nominal voltage of the cell \cite{Aaslid2020}; the Voltage-Current Model can be obtained from the Concentration-Current Model by matching the physical process inside the cell with the corresponding circuit component \cite{Berrueta2018}.   

\begin{figure}[t!]
	\begin{center}
		\includegraphics[trim=10cm 6cm 8cm 4cm, clip=true,angle=0,width=2.5in]{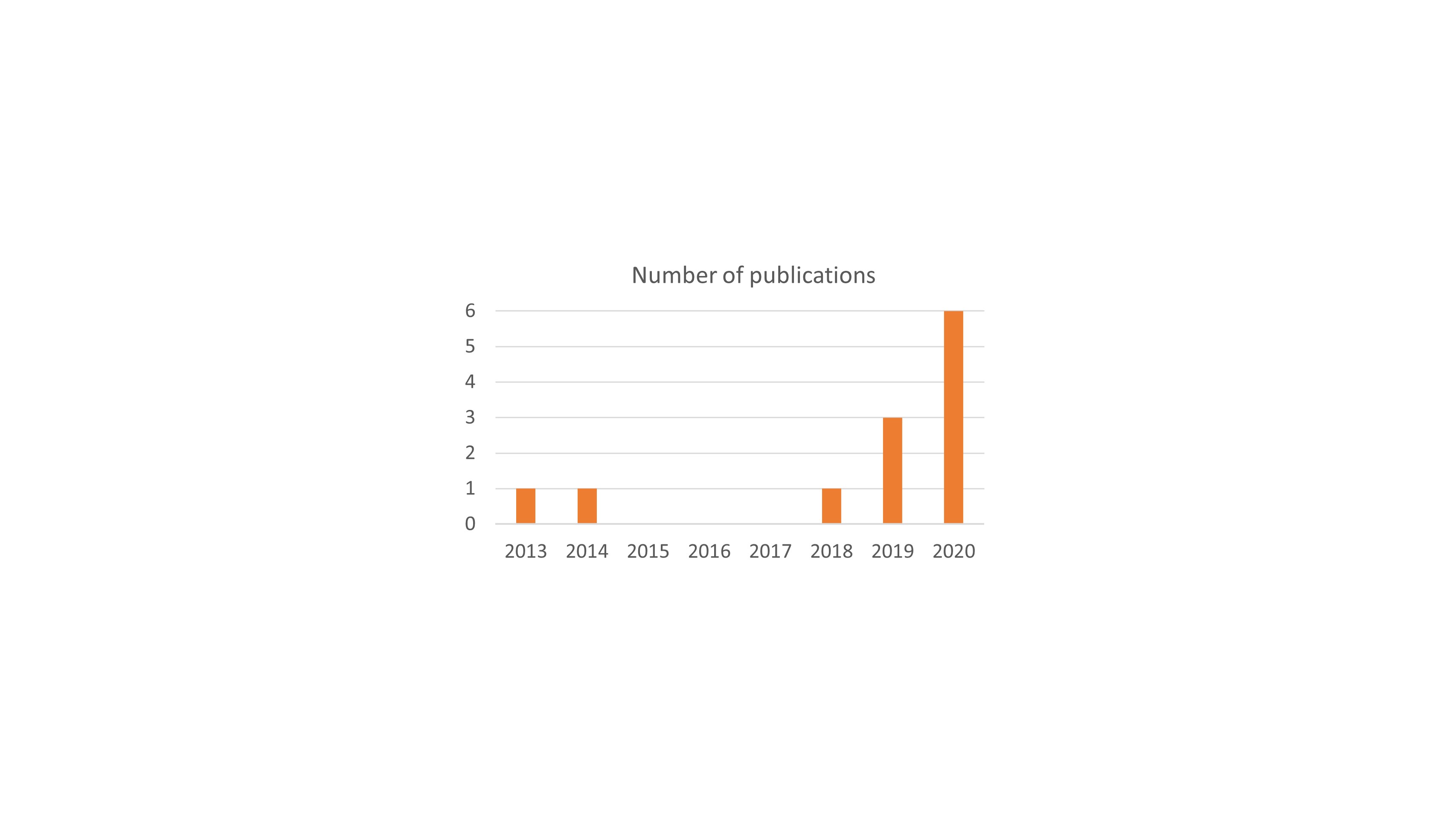}
		\caption[Number of publications in techno-economic studies of power system with the battery model different from a simple power-energy model.]{\label{fig:4} Number of publications in techno-economic studies of Power System with the battery model different from a simple power-energy model}
	\end{center}
\end{figure}

The interest to the models that describe the dynamics of the processes inside the battery has increased recently (Figure \ref{fig:4}). All the papers in power system economic studies employing either the Voltage-Current Model or the Concentration-Current Model were published in the last three years \cite{Reniers2018,Reniers2020,Cao2020, Taylor2020,Rosewater2019,Aaslid2020}. This can be linked with the appearance of the experimental data to benchmark the results of the ``optimal" schedule \cite{Reniers2020, Taylor2020}. Nonetheless, the Voltage-Current Model and the Concentration-Current Model are computationally expensive because of the number of constraints \cite{Rosewater2019} and a need for more time steps to improve stability \cite{Reniers2018}. 

\section{Alternative Battery Models in Power Systems Studies}\label{Alternative Battery Models in Power Systems Studies}

In this section, publications, where optimal charging/discharging schedules were identified for different BESS applications, are reviewed with the scope to define how BESS was modelled. The objectives of both system operators and independent storage owners are examined. The list of BESS applications is limited to system-level grid applications of BESS, namely: energy arbitrage, frequency regulation, operating reserve, peak shaving, renewable integration assistance, and transmission upgrade deferral. Papers that are used in this section are classified by the operation and degradation models in Table \ref{tab2} and Table \ref{tab3} respectively.

\begin{table}[t!]\centering
\caption{Literature survey on the battery grid applications (operation models).}
\begin{tabular}{@{}p{4cm}|p{3cm}|p{3cm}|p{3cm}@{}}
\hline
\textbf{Application} &  \textbf{Power-Energy Model} &\textbf{Voltage-Current Model} & \textbf{Concentration-Current Model} \\
\hline
 Energy arbitrage & \cite{Walawalkar2007,Lamont2013,Awad2014,Wankmuller2017,Fares2018,Maheshwari2020,Arcos-Vargas2020, He2018, Sakti2017,Pandzic2019,Gonzalez-Castellanos2020,He2020} & \cite{Reniers2018} &  \cite{Reniers2020,Reniers2018} \\
 Frequency regulation &  \cite{Zhang2018,Zhu2019,He2016,Xu2018,Shi2018,Sorourifar2020}&
    & 
    \cite{Cao2020} \\
 Operating reserve & \cite{Xu2018_Factoring,Padmanabhan2020,He2016} &   & 
   \\
 Peak shaving &  \cite{Schneider2020} &  \cite{Taylor2020} &  \cite{Rosewater2019} \\
Renewable integration assistance & \cite{Dicorato2012,Bhattacharjee2020,Shin2020,Jafari2020} &   &   \\
 Transmission upgrade deferral & \cite{Fernandez-Blanco2017,Falugi2018,Khani2016,Arteaga2021} &   &   \\
\hline
\end{tabular}
\label{tab2}
\end{table}

\begin{table}[t!]\centering
\caption{Literature survey on the battery grid applications (degradation models).}
\begin{tabular}{@{}p{4cm}|p{3cm}|p{3cm}|p{3cm}@{}}
\hline
\textbf{Application} &  \textbf{Without degradation} &\textbf{Empirical model} & \textbf{Physics-based model} \\
\hline
Energy arbitrage & \cite{Walawalkar2007,Lamont2013,Awad2014,Sakti2017,Pandzic2019,Gonzalez-Castellanos2020}  &   \cite{Wankmuller2017,Fares2018,Maheshwari2020,Arcos-Vargas2020, He2018,He2020} &  \cite{Reniers2018,Reniers2020} \\
 Frequency regulation & \cite{Zhang2018,Zhu2019} &  
 \cite{He2016,Xu2018,Shi2018,Sorourifar2020} &  \cite{Cao2020} \\
 Operating reserve & \cite{Nguyen2019} & 
  \cite{Xu2018_Factoring,Padmanabhan2020,He2016} &   \\
 Peak shaving  & \cite{Taylor2020}&  \cite{Schneider2020}  & \cite{Rosewater2019} \\
 Renewable integration assistance & \cite{Dicorato2012,Bhattacharjee2020} & \cite{Shin2020,Jafari2020} &   \\
Transmission upgrade deferral & \cite{Fernandez-Blanco2017,Falugi2018,Khani2016,Arteaga2021} &   &   \\
\hline
\end{tabular}
\label{tab3}
\end{table}

\subsection{Energy arbitrage}\label{Energy arbitrage}

Energy arbitrage is exploited by an independent BESS operator to generate revenue by charging the battery under the low-price conditions and by discharging when prices are higher. The economic analysis of the energy arbitrage application of BESS was accelerated by the restructuring and deregulation in the electric utility industry, and one of the first works on this topic was done by Graves \textit{et al.} \cite{Graves1999}.

Most works that estimate economic benefits of exploiting energy arbitrage by BESS employ a simple Power-Energy Model. In \cite{Walawalkar2007}, perfect information about price and demand and a price-taker model were used to assess the economic benefits of incorporating energy storage into the New York electricity market. The effect of large-scale energy storage on electricity price formation was examined in \cite{Lamont2013}. The strategic behavior of a BESS operator under price uncertainty in day-ahead and real-time electricity markets was addressed in \cite{Awad2014}. Several studies \cite{Wankmuller2017,Fares2018,Maheshwari2020,Arcos-Vargas2020, He2018} combined a Power-Energy Model with an empirical degradation model to estimate BESS profitability for energy arbitrage. The empirical based degradation formulation for the energy capacity with the assumption of linear dependence between energy throughput and extent of capacity fading significantly changed the cost-effectiveness of BESS investment \cite{Wankmuller2017}. In \cite{Fares2018}, the analysis of energy arbitrage is performed with so-called equivalent full cycles. In their market settings, they demonstrated that an extended calendar life is more profitable than an extended cycle life. The BESS operator would choose charging/discharging cycles with greater arbitrage trading profit if a longer calendar life is possible. Maheshwari and co-workers \cite{Maheshwari2020} applied their own experimental data with lithium-ion cells to derive their nonlinear degradation model. They claimed that DoD combined with cycle life and energy throughput quantification techniques for degradation fails to acknowledge the impact of state-of-energy and applied current, which were employed in their work through linear interpolation of their experimental data. In \cite{Arcos-Vargas2020}, the optimal value of maximum charging/discharging power was selected for the fixed capacity considering its degradation over the battery lifespan. Mohsenian-Rad \cite{Mohsenian-Rad2016} introduced charging and discharging bidding strategies in a stochastic framework for self-schedule and economic bids. Using short-term marginal cost per unit of degradation, which was derived from energy throughput and capital cost of the battery, He \textit{et al.} \cite{He2018} obtained a more accurate estimate of the energy arbitrage business case for the California day-ahead electricity market. Overall, the energy arbitrage operation considering ageing of the battery gives a better estimate in the cost/benefit analysis, but methods to characterize ageing are completely empirical and the results were not validated with real experience.

Some attempts have been made in order to improve the accuracy of the Power-Energy Model description for BESS representation used in energy arbitrage. It included the functional dependence between energy efficiency, state-of-energy, and maximum charging/discharging power. Sakti \textit{et al.} \cite{Sakti2017} updated a Power-Energy Model by considering the nonlinear dependence of the maximum charging/discharging power limits and energy efficiency on state-of-energy. Although the model was empirical by the formulation, authors affirmed that a simple Power-Energy Model of BESS may overestimate the earnings from energy arbitrage by 10\% compared with their most sophisticated model for a more volatile price signal resolution over the 7-day decision horizon. The authors of \cite{Pandzic2019} applied the results of their experimental findings to better define the limits of available charging power to reflect the constant-current/constant-voltage charging operation of the lithium-ion cell. The available charging power depends on the state-of-energy. If a generic Power-Energy Model is used for BESS characterization the optimal schedule in the energy arbitrage application may overestimate the profit by 300\% compared with the actual output obtained from scaled laboratory BESS that executed an ``optimal" schedule for their case study. Authors of \cite{Gonzalez-Castellanos2020} studied the optimal operation of BESS deployed in the IEEE-14 system to exploit arbitrage opportunities. Similar to \cite{Sakti2017}, their BESS model was an upgrade of a simple Power-Energy Model where fixed parameters were replaced with state-of-energy dependent ones. The use of nonlinear dependence between energy efficiency, charging/discharging power limits and state-of-energy was justified by the Voltage-Current Model as suggested by \cite{Berrueta2018}. Using empirical charging/discharging curves, authors concluded that for their case study a simplistic Power-Energy Model overestimated economic opportunities by 15\% and resulted in the operation of the battery beyond the recommended operating envelope. He \textit{et al.} \cite{He2020} combined a simple Power-Energy Model and energy throughput method for degradation description to derive the economic EoL of BESS. The term economic EoL was used by He to refer to the stage of the BESS state-of-health where the profit opportunities are vanished. In his optimization protocol energy efficiency, power, and energy capacity declined over time and as a result of cycling.
  
The energy arbitrage application is also a favourable choice to verify the robustness of more detailed operation models of BESS. Reniers \textit{et al.} \cite{Reniers2018} calculated economic benefits from energy arbitrage over one year of operation for three different battery models, i.e., a Power-Energy Model, a Voltage-Current Model and a Concentration-Current Model and three different degradation formulations. The authors compared these degradation descriptions with the experimental data and concluded that the concentration-current model was the most precise. They found that the energy arbitrage market participation strategy obtained for the Concentration-Current Model was considerably more profitable and with less degradation. In a more recent publication by the same authors \cite{Reniers2020} the optimal charging/discharging dispatch for energy arbitrage with Power-Energy Model and Concentration-Current Model were used to cycle lithium-ion cells in the laboratory conditions. The profit and the capacity loss were more accurately predicted by the Concentration-Current model. Moreover, physics-based approach for battery operation and ageing characterization reduced degradation by 30\% and improved revenue by 20\% compared with conventional Power-Energy Model with empirical degradation.

\subsection{Frequency regulation}\label{Frequency regulation}

Frequency regulation is one of the ancillary service products and it is needed to keep frequency within an acceptable range when there is a mismatch between supply and demand. The fast ramping capabilities of BESS make it a favourable choice for frequency regulation: for example, 75\% of large-scale BESS power capacity in US is used for the balancing of momentary fluctuations in the system \cite{US_energy_2020}. The frequency regulation application of BESS in power system economics studies is rarely examined solely and it is usually combined with energy arbitrage.

The BESS is mostly modelled through a generic Power-Energy Model \cite{Zhang2018, Zhu2019} for charging/discharging performance whereas the degradation is characterized by means of an empirical relationship \cite{He2016,Xu2018,Shi2018,Sorourifar2020}. In \cite{Zhang2018}, the authors investigated two-level, planning and operation, strategy for BESS owner to maximize profit from the frequency regulation market. Zhu \textit{et al.} \cite{Zhu2019} derived strategic operation for an aggregator coordinated BESS. Xu \cite{Xu2018} found the optimal control policy for BESS deployed in their case study to boost profit from the performance-based frequency regulation market using a chance-constraint optimization. The author used the cycle counting algorithm to characterize the long-term performance.  The optimal bidding strategy in the frequency regulation market for the electric vehicle aggregator with different participation scenarios was outlined in \cite{Vagropoulos2013}. Their model was able to simulate the transition from constant current mode to constant voltage mode of charging operation. They  highlighted that their representation of the lithium-ion battery resulted in a more accurate estimate of financial benefits: there was up to 20\% difference compared with a generic Power-Energy Model associated with BESS. Shi and co-workers \cite{Shi2018} coupled a Power-Energy Model with three different degradation models, namely, the fixed cost of degradation, the energy throughput method, and cycle-based model based on the rainflow algorithm, to explore the financial benefits of frequency regulation for the BESS owner in the their case study. It was shown that the rainflow cycle-based degradation model projects up to 27.6\% growth in revenue, and thus a greater return on investment, and almost 85\% increase in battery life expectancy. In \cite{Sorourifar2020}, a Power-Energy Model coupled with the energy throughput ageing quantification technique was used to find the optimal energy capacity of BESS, which profits from energy arbitrage and frequency regulation with compensation for capacity and energy provided in the California day-ahead and real-time energy and ancillary services markets. Assuming a constant degradation rate, the loss of capacity was directly incorporated into the operation framework as a constraint.

In \cite{Cao2020}, a Concentration-Current Model was employed to find the optimal schedule of BESS for the frequency regulation market. The strategy was compared with one obtained using a Power-Energy Model with degradation cost, which was proportional to capacity committed to frequency regulation, in the objective function and with one where degradation was calculated using the SEI method in postprocessing of the optimal schedule. The authors reported a 143\% increase in lifetime and 35\% growth in profit compared to a generic Power-Energy Model.

\subsection{Operating reserve}\label{Operating reserve}

The operating reserve is intended for the grid frequency management if a significant unpredictable deviation to the supply/demand balance occurs in the system. The operating reserve revenue stream for the BESS owner is usually considered as an additional source of revenue and it is bundled with other BESS applications in power system economic analysis \cite{Arteaga2019}.

The strategic operation of BESS that provides operating reserve services is usually derived with a simple Power-Energy Model bundled with empirical ageing models \cite{Xu2018_Factoring,Padmanabhan2020,He2016}. Xu \textit{et al.} \cite{Xu2018_Factoring} proposed a dispatch strategy for BESS considering the cost of battery degradation that was formulated using the Rainflow cycle counting algorithm. In \cite{Padmanabhan2020}, the optimal strategy for BESS operator that submits the bids into both the energy and operating reserve market was derived by combining the impact of DoD and the discharge rate in the degradation cost function. Reference \cite{He2016} modelled optimal market participation of BESS in three day-ahead markets: energy, frequency regulation and operating reserve markets. The profit of BESS was calculated on a daily basis and prorated by the number of the battery's daily equivalent 100\%-DoD cycles. Nguyen \textit{et al.} \cite{Nguyen2019} replaced the fixed energy efficiency of a generic Power-Energy Model with one that nonlinearly depends on the charging/discharging power and state-of-energy. The parameters for their empirical model were calculated from the charging/discharging curve provided by the cell manufacturer. The proposed nonlinear model estimated almost 17\% less of the total revenue over one year compared with simple Power-Energy Model if it was installed in their market environment. Perez \textit{et al.} \cite{Perez2016} examined the impact of applying practical box constraints on state-of-energy to limit degradation on the financial potential of BESS providing several services including OR. Although there was a drop in revenue from energy arbitrage, the net revenue has increased from operating reserve and frequency regulation contributions because of extended lifespan.

\subsection{Peak shaving}\label{Peak shaving}

A BESS can replace the need to construct a new peaking generation capacity that would meet the peak demand from a highly volatile load. If a BESS is installed on the load side it could also work as a peak shaver to minimize the total electricity bill. There are several works where charging/discharging decisions were found for the peak shaving application.

Schneider \textit{et al.} \cite{Schneider2020} investigated a strategic investment into BESS for a bundled peak shaving and energy arbitrage business model. Their optimization framework included Power-Energy Model, Rainflow cycle counting paradigm for battery degradation and it was solved heuristically through three stages: on the first stage, the daily scheduling maximizes energy arbitrage revenue corrected by a degradation penalty term, on the second stage the total monthly revenue from peak shaving and energy arbitrage is calculated, the third stage is used to define the optimal schedule over the year. Taylor \cite{Taylor2020} employed Voltage-Current Model formulation for BESS and compared it  with Power-Energy Model. The parameters of the Voltage-Current Model were measured from their own lab experiments with lithium iron phosphate battery cells. First, the accuracy of the model was demonstrated by comparison with the experiment. Second, the optimal schedule of BESS based on the Voltage-Current Model formulation outperformed the results with Power-Energy Model when the optimal schedule of each model was processed by the battery hardware simulation tool. Although simulation was carried out at only two fixed levels of the current rate, it was clear that a generic Power-Energy Model model did not ensure reliable performance. In the review of battery models for the optimal control \cite{Rosewater2019}, the control strategy for BESS installed to reduce the total electricity bill over 24-hour decision horizon was obtained for three different battery models, namely a Power-Energy Model, a Voltage-Current Model and the Concentration-Current Model. The cost reduction in the bill was conducted using energy arbitrage and peak shaving applications of BESS. The degradation of the lithium-ion cell was not a part of the analysis. Although the net reduction in the electricity bill was almost the same for all models and stood at about 8\%, the authors claimed that the control strategy for the Power-Energy Model was likely infeasible. The Voltage-Current Model and Concentration-Current Model gave similar results for the state variables such as voltage and current.

\subsection{Renewable integration assistance}\label{Renewable integration assistance}

BESS can also accelerate the integration of intermittent renewable capacity to the grid by mitigating its natural fluctuation and can increase the return on investment in a renewable generation project if it is combined with a BESS \cite{US_energy_2020}. The objective of the optimization problems with this application can be either finding the optimal operation schedule or the planning problem where the optimized BESS size is explored to maximize return on investment over the projected lifespan of a battery. 

Similar to operation studies with other BESS applications, the renewable integration assistance application of BESS is mostly modelled with a simple Power-Energy Model \cite{Dicorato2012,Bhattacharjee2020}. In \cite{Dicorato2012} the size and optimal dispatch were determined for BESS paired with a wind farm. The solution enhanced the operational stability and economic feasibility of the wind power project. Bhattacharjee \textit{et al.} \cite{Bhattacharjee2020} used a generic Power-Energy Model to optimally size energy storage and transmission interconnector, which were coupled with a wind power facility, for the strategic participation in the energy market. Recently more works \cite{Shin2020,Jafari2020} have been presented with BESS models that can ensure reliable performance and characterize capacity and power fading.  Using a heuristic algorithm Shin \cite{Shin2020} investigated the impact of BESS size on degradation while searching for the optimal BESS capacity supplementing photovoltaic generation for two scenarios of battery use: constant usable energy capacity and a fixed DoD. Jafari \cite{Jafari2020} studied the economic impact of pairing offshore wind farm with BESS. This model of BESS included varying energy efficiency and power limits. Calendar and cycling ageing of capacity were also incorporated in the model by a linear decline assumption for calendar ageing and combination of the energy throughput with the number of equivalent full cycles for cycling ageing respectively. The revenue from the optimal schedule with enhanced Power-Energy Model without degradation was 4\% less than for the schedule obtained with a simplistic Power-Energy Model. When one of the degradation models was added to the optimization framework the estimated revenue decreased by 35\%. The Voltage-Current model was employed in \cite{Aaslid2020} to determine the optimal schedule of BESS over 36-hour optimization horizon while minimizing the electricity bill of the user with on-site photovoltaic generation. The authors showed that this model avoids unsafe operation compared to the power-energy model.

\subsection{Transmission upgrade deferral}\label{Transmission upgrade deferral}

Another BESS application that brings significant changes to how the long-term planning of the power system is performed, is a deferral or replacement of the traditional power system infrastructure – transmission lines. When BESS is installed downstream of a congested transmission corridor it can relieve congestion by discharging to meet the additional demand from the load. This is why, BESS is considered as a virtual transmission for the future grid. The strategic deployment of BESS can increase the asset utilization rate in the grid if it is planned correctly \cite{Pandzic2015}. 

The class of planning problems from the system operator perspective, where the optimal size of BESS and its location in the grid are determined, is usually solved with the objective to minimize the sum of the capital cost of BESS and the operating costs of the system. A simple Power-Energy Model without degradation is usually incorporated into these optimization problems. For example, in \cite{Fernandez-Blanco2017}, a static investment model was used to find siting and sizing decisions within the Western Electricity Coordinating Council interconnection by means of the stochastic programming. Falugi and co-workers \cite{Falugi2018} studied the dynamic planning problem of the joint transmission and BESS deployment in the IEEE 118-bus system for the planning horizon of 16 years. The optimal decision plan was updated every four years.

The BESS that provides transmission services should be paid through the rate-based compensation. However, if BESS also provides other services to the grid it should be additionally compensated through the market. Such multiple service operation of BESS and corresponding compensation scheme currently face several regulatory barriers. The market model and corresponding policies for storage as a transmission asset are investigated by several utilities \cite{CAISO_SATA_2018}. The operation strategy for energy storage that provided congestion relief service and also obtained revenue from energy arbitrage is examined in \cite{Khani2016}.  The optimal premium paid to the BESS owner as a rate-based compensation to relieve congestion is explored in \cite{Arteaga2021}. Both papers utilized a generic Power-Energy Model without degradation.

\section{Discussion}\label{Discussion}

The system-level operational and planning studies predominantly employed generic Power-Energy Models to characterize the BESS charging/discharging performance and various empirical models to quantify degradation. The reason for this is the simplicity and linearity of Power-Energy Models. From the reviewed literature it can be concluded that advanced battery models can provide more accurate estimates for the economic potential of BESS, feasible charging/discharging schedule, and more precise projection of the capacity and charging/discharging power fading.

Several authors \cite{Sakti2017,Pandzic2019, Vagropoulos2013,Gonzalez-Castellanos2020} enhance a simplistic Power-Energy Model with the functional dependences between energy efficiency, maximum charging/discharging power and state-of-energy to better model typical characteristics of the lithium-ion cell. The linear approximation was applied for all mentioned relationships to make them solvable for the optimization problems used in those studies. However, only in the case of \cite{Sakti2017} the final problem was a mix-integer linear problem whereas authors of \cite{Pandzic2019, Vagropoulos2013, Gonzalez-Castellanos2020} finished with a linear programming problem. The optimal schedule for BESS with a simplistic Power-Energy Model, in general, overestimated economic valuations. 

The energy arbitrage application was used for the assessment of BESS models from \cite{Sakti2017, Pandzic2019,Gonzalez-Castellanos2020} whereas the frequency regulation service revenue stream was assessed in \cite{Vagropoulos2013}. The common drawback for these models is that they are phenomenological by their nature: limited experimental data were used to fit their mathematical models for selected operating conditions. Moreover, all models were run over the narrow optimization horizon of one to seven days. The degradation of BESS was not considered as only charging/discharging performance was an objective for the improvement. In \cite{Jafari2020}, the strategic sizing of BESS with renewable generation was not part of the problem, but they claimed a simplified model could exaggerate the revenue by 35\%. This is a significant error in the evaluation of the economic feasibility  and may lead to misleading conclusions in \cite{Bhattacharjee2020} where a simplistic model of BESS for planning studies was used.

An increasing number of studies for different BESS applications such as \cite{Maheshwari2020} for energy arbitrage, \cite{Schneider2020} for peak shaving, \cite{Xu2018} for frequency regulation, and \cite{Padmanabhan2020} for operating reserve showed that the economic viability of the project with BESS is not overestimated if the degradation models were used. However, various empirical degradation formulations, which are limited by their formulation, were used in the mentioned works. 

Among three models to simulate the charging/discharging profile, the Concentration-Current Model can be seen as the most promising since it characterizes the dynamics of physical processes inside the cell and can be coupled with the physics-based degradation model such as SEI formation. The cost-benefit analysis of BESS with Concentration-Current Model was performed only for a discrete battery application such as energy arbitrage in \cite{Reniers2018}, frequency regulation \cite{Cao2020}, and peak shaving \cite{Rosewater2019}.  The co-optimization of various BESS applications was not considered when this model is employed. This high-fidelity model was also not used for optimal sizing of BESS for the transmission services or renewable integration assistance. As the optimization problem with Concentration-Current Model is computationally expensive, authors of \cite{Rosewater2019} only considered a 24 hour interval for their case study whereas authors of \cite{Reniers2018} and \cite{Cao2020} utilized the model predictive control scheme for 24 hours and 1 hour respectively.

Based on these observations, several directions are suggested for future development:

\begin{itemize}
	\item The impact of the detailed model of BESS without degradation on the profit-maximizing operation of BESS for various grid applications was quantified by many researchers. However, there is no consensus on how a more detailed model is crucial for short-term operation. For example, in \cite{Pandzic2019}, the high-level model overestimated the profit by 20\%, 4\% was demonstrated in \cite{Jafari2020}, and no difference was stated by \cite{Reniers2020}. 
	\item An economic feasibility for only energy arbitrage, frequency regulation, and peak shaving was done with the concentration-current models. The analysis for the case of other services or several stacked application of BESS was out of consideration.
	\item The physics-based model has never been considered for planning studies as the system-level planning studies predominantly employed a generic Power-Energy model to characterize the BESS charging/discharging performance. As the lifespan of the lithium-ion cell component of a BESS is a quarter or half of traditional transmission and generation assets, the integration of BESS into the grid requires a multistage planning approach, where a replacement schedule is a part of the implementation plan and investment. The long-term multistage battery planning with replacement is completely out of consideration in the literature.
	\item The increased number of variables and constraints in the optimization framework brought by voltage-current and concentration-current  models  should be tackled with parallel computing as it was performed for a longer decision-making horizon with a Power-Energy model considering degradation in \cite{Sorourifar2020}.
	\item The stochastic formulation of the strategic operation of BESS is a computationally expensive problem by itself. The need in using a more detailed battery model for this optimization framework should be justified.
		
	\item A highly efficient, time-saving algorithm is needed for the nonlinear optimization problem of the planning and scheduling of the grid level applications if a sophisticated physics-based lithium-ion model is employed.  
\end{itemize}

\section{Conclusion}\label{Conclusion}

In this paper, three BESS models for operation with different level of details were reviewed and their governing equations that are appropriate for the optimization framework are presented. Several degradation models were discussed and their differences were highlighted. The comprehensive literature review of research papers where optimal operation and planning decisions were derived for the business cases with battery storage for various system-level applications was performed. This research review showed that a more sophisticated battery description ensures more accurate estimates of BESS economic benefits, longer projected lifespan and operation within safety limits. Currently, the number of research papers with the detailed physics-based model of a BESS is limited. The integration of such models in a broader spectrum of problems with the strategic operation and planning presents an attractive direction for the future research. 

\bibliographystyle{model1-num-names.bst}
\bibliography{References_PhD_all}

\end{document}